# The spin Nernst effect in Tungsten

Peng Sheng[1], Yuya Sakuraba[1], Yong-Chang Lau[1,2], Saburo Takahashi[3], Seiji Mitani[1] and Masamitsu Hayashi[1,2*]

[1]*National Institute for Materials Science, Tsukuba 305-0047, Japan*

[2]*Department of Physics, The University of Tokyo, Bunkyo, Tokyo 113-0033, Japan*

[3]*Institute for Materials Research, Tohoku University, Sendai 980-8577, Japan*

**The spin Hall effect allows generation of spin current when charge current is passed along materials with large spin orbit coupling. It has been recently predicted that heat current in a non-magnetic metal can be converted into spin current via a process referred to as the spin Nernst effect. Here we report the observation of the spin Nernst effect in W. In W/CoFeB/MgO heterostructures, we find changes in the longitudinal and transverse voltages with magnetic field when temperature gradient is applied across the film. The field-dependence of the voltage resembles that of the spin Hall magnetoresistance. A comparison of the temperature gradient induced voltage and the spin Hall magnetoresistance allows direct estimation of the spin Nernst angle. We find the spin Nernst angle of W to be similar in magnitude but opposite in sign with its spin Hall angle. Interestingly, under an open circuit condition, such sign difference results in spin current generation larger than otherwise. These results highlight the distinct characteristics of the spin Nernst and spin Hall effects, providing pathways to explore materials with unique band structures that may generate large spin current with high efficiency.**

*Email: hayashi@phys.s.u-tokyo.ac.jp

## INTRODUCTION

The giant spin Hall effect(*1*) (SHE) in heavy metals (HM) with large spin orbit coupling has attracted great interest owing to its potential use as a spin current source to manipulate magnetization of magnetic layers(*2-4*). Recently it has been reported(*5, 6*) that the spin Hall conductivity of *5d* transition metals depends on the number of *5d* electrons, indicating that the observed SHE is due to the topology and filling of the characteristic bands at the Fermi surface(*7, 8*). Spin current in solids can be produced not only by charge current but also by heat current(*9*). Understanding the coupling between spin current and heat current is the central subject of Spin caloritronics(*10*). It is now well understood that a temperature gradient applied across a magnetic material, typically a magnetic insulator, results in spin accumulation that can be used to generate spin current in neighboring non-magnetic materials via the spin Seebeck effect(*11-17*).

It has been predicted theoretically(*18-23*) that in non-magnetic materials with strong spin orbit coupling, the heat current can be converted into spin current. The effect, often referred to as the spin Nernst effect, generates spin current that scales with the energy derivative of the spin Hall conductivity. Here we show direct probe of the spin Nernst effect in amorphous-like W, which possesses the largest spin Hall angle among the *5d* transition metals(*6, 24, 25*). When an in-plane temperature gradient is applied across W/CoFeB/MgO heterostructures, we observe longitudinal and transverse voltages that vary with the magnetic field similar to those of the spin Hall magnetoresistance(*26-31*). The W layer thickness dependence of the longitudinal voltage is compared to that of the spin Hall magnetoresistance to estimate the size and sign of the spin Nernst angle. We find the spin Nernst angle of W is slighter smaller than (~70%) its spin Hall angle and the two angles possess opposite sign.

**RESULTS**

The film structure used is sub.|$d_N$ HM|1 FM|2 MgO|1 Ta (thickness in units of nanometer), where HM is Ta or W and the ferromagnetic metal (FM) is $Co_{20}Fe_{60}B_{20}$ (referred to as CoFeB hereafter). We first study the electrical transport properties of the films. The inverse of the device longitudinal resistance ($1/R_{XX}$) multiplied by a geometrical factor ($L/w$), i.e. the sheet conductance ($G_{XX}$), is plotted as a function of the HM layer thickness ($d_N$) in Figs. 1(a) and 1(b) for the Ta and W underlayer films, respectively. (See inset of Fig. 1(a) for the definitions of $L$ and $w$ as well as the coordinate system.) We fit the data with a linear function to estimate the resistivity ($\rho_N$) of the HM layer. The fitting results are shown by the blue solid lines: we obtain $\rho_N$~183 μΩcm for Ta and ~130 μΩcm for W. Note that W undergoes a structural phase transition(*6, 24, 32*) when its thickness is larger than ~6 nm, as indicated by the change in $G_{XX}=L/(wR_{XX})$ at this thickness. The spin Hall magnetoresistance (SMR), $R_{SMR}=\Delta R_{XX}/R_{XX}^Z$, is plotted as a function of the HM layer thickness in Figs. 1(c) and 1(d). We define $\Delta R_{XX}$ as the resistance difference when the magnetization of the CoFeB layer is pointing along $y$ ($R_{XX}^Y$) and $z$ ($R_{XX}^Z$) directions, i.e. $\Delta R_{XX}=R_{XX}^Y-R_{XX}^Z$. The thickness dependence of $R_{SMR}$ is consistent with previous reports(*6, 31*).

The transverse resistance of the films is shown in Fig. 2 (see the inset of Fig. 1(a) for the details of the measurement setup). The inset of Fig. 2(a) shows the transverse resistance ($R_{XY}$) vs. the out of plane field $H_Z$ for a Ta underlayer film. We define $2\Delta R_{XY}$, i.e. the anomalous Hall resistance, as the difference in $R_{XY}$ when the magnetization is pointing along $+z$ and $-z$. In Figs. 2(a) and 2(b), $\Delta R_{XY}$ is plotted as a function of HM layer thickness. $\Delta R_{XY}$ decreases with increasing $d_N$ largely due to current shunting into the HM layer. To estimate the anomalous Hall angle, $\Delta R_{XY}$ is divided by $R_{XX}^Z$, multiplied by a geometric factor ($L/w$), and divided by a constant ($x_F$) that accounts for the current shunting effect into the HM layer: $x_F = \dfrac{t_F \rho_N}{t_F \rho_N + d_N \rho_F}$, where $t_F$ and $\rho_F$ are the thickness and resistivity of the FM layer, respectively. The HM layer thickness dependence of the normalized anomalous Hall coefficient $R_{AHE}/x_F = (\Delta R_{XY} L)/(R_{XX}^Z w x_F)$ is plotted in Figs. 2(c) and 2(d) for Ta and W underlayer films. We find that the normalized anomalous Hall coefficient shows a significant HM layer thickness dependence, in particular, for the W underlayer films.

We next show the thermoelectric properties of the films. Figure 3(a) shows a sketch of the setup to study the Seebeck coefficient of the films. A heater is placed near one side of the substrate to create a temperature gradient across the substrate. The difference in the temperature between the hot ($T_H$) and cold ($T_L$) sides of the substrate ($\Delta T = T_H - T_L$), across a distance $D$, is measured using an infrared camera. The longitudinal (Seebeck) voltage $V_{XX} = V(x_1) - V(x_2)$ is measured between two points of the device separated by a distance $L$ ($L<D$). The temperature of position $x_1$ is higher than that of $x_2$; see Fig. 3(a). The $\Delta T$ dependence of $V_{XX}$ is shown in Figs. 3(b) and 3(c) for the Ta and W underlayer films, respectively. The data is fitted with a linear function to extract the Seebeck coefficient(33) $S = -(V_{XX}/L)/(\Delta T/D)$ from the slope, which is plotted as a function of $d_N$ in Figs. 3(d) and 3(e). $S$ approaches $\sim−4$ µV/K when the HM layer thickness is thin for both film structures, which we consider provides information of the Seebeck coefficient of CoFeB (we assume that the MgO and the oxidized Ta capping layers have negligible contribution to $V_{XX}$). In contrast, the thick limit of $d_N$ gives the Seebeck coefficient of the HM layer: we estimate $S\sim−2$ µV/K for Ta and $\sim−12$ µV/K for W.

The off diagonal component of the thermoelectric properties is summarized in Fig. 4. The experimental setup to study the temperature gradient induced transverse voltage is depicted in Fig. 4(a). A typical hysteresis loop obtained by measuring the $H_Z$ dependence of the temperature gradient induced transverse voltage $V_{XY} = V_{XY}(y_2) − V_{XY}(y_1)$ (see Fig. 4(a) for the definitions of $y_1$ and $y_2$) is shown in Fig. 4(b). Similar to the anomalous Hall resistance, we define $2\Delta V_{XY}$, i.e. the anomalous Nernst voltage, as the difference in $V_{XY}$ when the magnetization is pointing along $+z$ and $−z$. Figure 4(c) shows the $\Delta T$ dependence of $\Delta V_{XY}$ for a W underlayer film. Within the applied temperature gradient, the response is linear. We thus fit a linear function to obtain the anomalous Nernst coefficient $S_{ANE} = (\Delta V_{XY}/L)/(\Delta T/D)$ from the slope (here $L = y_2 − y_1$).

The HM layer thickness dependence of anomalous Nernst coefficient $S_{ANE}$ is plotted for the Ta and W underlayer films in Figs. 4(d) and 4(e), respectively. $|S_{ANE}|$ decreases with increasing $d_N$ for the Ta underlayer films whereas it shows a peak at around $d_N\sim3$ nm for the W underlayer films. Similar to the anomalous Hall resistance, the presence of the HM layer can shunt the Hall voltage. To account for such effect, $S_{ANE}$ is divided by $x_F$. The normalized anomalous Nernst coefficient $S_{ANE}/x_F$ is plotted as a function of $d_N$ in Figs. 4(f) and 4(g). We find a larger variation of $S_{ANE}/x_F$ with $d_N$ for the W underlayer films than the Ta underlayer films.

Recent studies have shown that spin current generated within the HM layer modifies the anomalous Hall resistance via a non-zero imaginary part of the spin mixing conductance at the HM/FM interface(*26, 27, 34*). The large variation of the normalized anomalous Nernst coefficient with $d_N$ for the W underlayer films indicates that a temperature gradient can cause spin current generation in the W layer that results in modification of the off diagonal component. To evaluate the temperature gradient induced spin current generation, i.e. the spin Nernst effect, in a more explicit way, we have studied the external field dependence of the Seebeck voltage in analogy to the SMR. The experimental setup is the same with that of Fig. 3(a): here a large external magnetic field is applied during the measurements.

In Figs. 5(a) and 5(b), we show the longitudinal (Seebeck) voltage $V_{XX}=V(x_1)-V(x_2)$ of Ta and W underlayer films, respectively, plotted as a function of external field directed along the y-axis ($H_Y$). The temperature difference $\Delta T$ across the substrate is ~3.5 K. For the W underlayer films (Fig. 5(b)), we find a peak-like structure around zero field. (Signals are shifted vertically for clarity so that the large field limit of $V_{XX}$ equals zero.) The peak found in the $V_{XX}$ vs. $H_Y$ plot decays to zero when $|H_Y|\sim|H_K|$, where $H_K$ is the effective anisotropy field required to force the magnetization to point along the film plane (see supplementary materials for the magnetic properties of the heterostructures). The peak amplitude $\Delta V_{XX}$ defined schematically in Fig. 5(b) is equivalent to the difference in $V_{XX}$ when the magnetization is pointing along the y-axis ($V_{XX}^Y$) and the z-axis ($V_{XX}^Z$), i.e. $\Delta V_{XX}=V_{XX}^Y-V_{XX}^Z$. Such definition is in accordance to that of SMR. We have also studied $V_{XX}$ as a function of $H_X$ and $H_Z$: the results are shown in the supplementary material (Fig. S2). In contrast to $V_{XX}$ vs. $H_Y$, we find no clear feature in the $H_X$ and $H_Z$ dependence of $V_{XX}$. These results suggest that the thermal analogue of the anisotropic magnetoresistance (AMR) is small in CoFeB(*35*). Note that the AMR of the CoFeB layer here is ~0.1%(*31*), much smaller than that of the Ni-based soft magnetic materials(*36*). The small temperature gradient induced AMR-like voltage ($V_{XX}$ vs. $H_X$, see Fig. S2) found here also indicates that possible contribution from combination of AMR and interfacial spin orbit coupling(*37, 38*) on $\Delta V_{XX}$ may be small. We also find little of evidence of proximity induced magnetism(*39-41*) in W and Ta, which may influence the temperature gradient induced voltage via AMR in the HM layer.

In Fig. 5(c), we plot $S_{SNE}=(\Delta V_{XX}/L)/(\Delta T/D)$, which we refer to as the spin Nernst coefficient, as a function of the W layer thickness. Interestingly $|S_{SNE}|$ takes a maximum at $d_N$~3-4 nm, similar to that of the SMR shown in Fig. 1(d). These results indicate that the interfacial magnetoresistance caused by the Rashba interaction, which takes a maximum at a HM layer thickness close to one lattice constant(*42*), is not the main source of the voltage ($S_{SNE}$) found here. See supplementary material (Figs. S3 and S4) for discussions on the effects of the FM layer (CoFeB) and an unintended out of plane temperature gradient(*15, 43-45*) on the voltage measurements.

To account for these results, a drift-diffusion model is extended to describe spin transport in a bilayer system. The HM layer thickness dependence of the SMR and the anomalous Hall coefficient are described with the following equations(*26, 31*):

$$R_{SMR} \equiv \frac{\Delta R_{XX}}{R_{XX}^Z} = -(1-x_F)\theta_{SH}^2 \frac{\lambda_N}{d_N}\tanh^2(\frac{d_N}{2\lambda_N})\text{Re}\left[\frac{g_S}{1+g_S\coth(d_N/\lambda_N)}\right] \quad (1)$$

$$R_{\text{AHE}} \equiv \frac{\Delta R_{XY}}{R_{XX}{}^Z} \frac{L}{w} = -x_F \theta_{AH} + (1-x_F)\theta_{SH}^2 \frac{\lambda_N}{d_N} \tanh^2(\frac{d_N}{2\lambda_N}) \text{Im}\left[\frac{g_S}{1+g_S \coth(d_N/\lambda_N)}\right] \quad (2)$$

where $\theta_{SH}$ and $\lambda_N$ are the spin Hall angle and the spin diffusion length of the HM layer, $\theta_{AH}$ is the anomalous Hall angle of the FM layer. $g_S = 2\rho_N \lambda_N G_{MIX}$, where $G_{MIX}$ is the spin mixing conductance of the HM/FM interface. Here for simplicity we have neglected contribution of longitudinal spin current absorption on the SMR(*31*).

Furthermore, we assume that a temperature gradient ($\nabla T$) applied across a sample can generate spin current $Q$ (i.e. flow of spin-angular momentum carried by electrons) via the spin Nernst effect in a similar way an electric field $E$ (or current) generates spin current through the spin Hall effect, i.e.

$$Q_{kj} = \frac{\hbar}{2|e|}\theta_{SH}\left(e_k \times \frac{E}{\rho_N}\right)\bigg|_j + \frac{\hbar}{2|e|}\theta_{SN}\left(e_k \times \frac{S_N}{\rho_N}(-\nabla T)\right)\bigg|_j \quad (3)$$

Indices $k$ and $j$ denote, respectively, the spin and flow direction of the spin current ($e_k$ is an unit vector.) $\hbar$ is the reduced Planck constant, $e$ is the electron's charge. $S_N$ and $\theta_{SN}$ are the Seebeck coefficient and the spin Nernst angle of the HM layer, respectively. For simplicity, we do not consider the spin Hall and spin Nernst effects of the FM layer since $\theta_{SH}$ of FM has been reported to be small compared to that of the HM layers(*46-48*). In the FM layer, however, the anomalous Hall and the anomalous Nernst effects generate a transverse charge current $J^T$ when $E$ and $\nabla T$ are applied. The transverse charge current (opposite to the electron flow) is:

$$J_j^T = -\theta_{AH}\left(m \times \frac{E}{\rho_F}\right)\bigg|_j - \theta_{AN}\left(m \times \frac{S_F}{\rho_F}(-\nabla T)\right)\bigg|_j \quad (4)$$

where $\hat{m}$ is an unit vector representing the magnetization direction of the FM layer. $S_F$ and $\theta_{AN}$ are the Seebeck coefficient and the anomalous Nernst angle of the FM layer, respectively.

We assume a temperature gradient $\frac{\Delta T}{D} = \frac{T_H - T_L}{D}$ is applied under an open circuit condition. The change in the longitudinal voltage ($\frac{V_{XX}}{L} = \frac{V_{XX}(x_1) - V_{XX}(x_2)}{x_2 - x_1}$) when the magnetization of the FM layer is pointing along $y$ ($V_{XX}^Y$) and $z$ ($V_{XX}^Z$) axes, $\Delta V_{XX} = V_{XX}^Y - V_{XX}^Z$, is expressed as:

$$S_{\text{SNE}} \equiv \frac{\Delta V_{XX}/L}{\Delta T/D} = (1 - x_F) \theta_{SH} \{\theta_{SH} S - \theta_{SN} S_N\} \frac{\lambda_N}{d_N} \tanh^2(\frac{d_N}{2\lambda_N}) \text{Re}\left[\frac{g_S}{1 + g_S \coth(d_N/\lambda_N)}\right] \quad (5)$$

Similarly, the difference in the transverse voltage ($\frac{V_{XY}}{L} = \frac{V_{XY}(y_2) - V_{XY}(y_1)}{y_2 - y_1}$) when the magnetization reverses its direction from $+z$ to $-z$, $2\Delta V_{XY} = V_{XY}^Z - V_{XY}^{-Z}$, reads:

$$\begin{aligned} S_{\text{ANE}} &\equiv \frac{\Delta V_{XY}/L}{\Delta T/D} = x_F \{\theta_{AH} S - \theta_{AN} S_F\} \\ &\quad - (1 - x_F) \theta_{SH} \{\theta_{SH} S - \theta_{SN} S_N\} \frac{\lambda_N}{d_N} \tanh^2(\frac{d_N}{2\lambda_N}) \text{Im}\left[\frac{g_S}{1 + g_S \coth(d_N/\lambda_N)}\right] \end{aligned} \quad (6)$$

Equations (5) and (6) represent the $d_N$ dependence of the spin Nernst and anomalous Nernst coefficients, respectively. The Seebeck coefficient of the HM/FM bilayer, defined as $S = x_F S_F + (1 - x_F) S_N$, is obtained experimentally using the relation $S = -(V_{XX}/L)/(\Delta T/D)$ and the results are shown in Figs. 3(d,e). We note that when $\theta_{SN}=0$, $S_{SNE} = S\,R_{SMR}$: the functional form of $S_{SNE}$ and $R_{SMR}$ is the same.

The first term ($\theta_{SH}S$) in the curly bracket of Eq. (5) appears due to the open circuit condition. That is, the electrons initially move from the hot to cold side when a temperature gradient is applied (the Seebeck coefficients of the FM and HM layers are all negative). Once the electrons reach the edge of the patterned structure, an internal electric field $\boldsymbol{E}_{INT}$ develops due to charge accumulation at the edges. The direction of $\boldsymbol{E}_{INT}$ is such that it cancels the electron flow driven by the temperature gradient, resulting in a net zero current. However, spin current can be generated via the spin Hall effect when a non-zero $\boldsymbol{E}_{INT}$ exists, thus contributing to the SMR. The second term ($\theta_{SN}S_N$) in the curly bracket of Eq. (5) corresponds to contribution to the SMR that results from a direct conversion of heat current to spin current. Similar classification also applies to the terms in the curly brackets of Eq. (6).

The model calculations are compared to the experimental results presented in Figs. 1(c,d), 2(c,d), 4(f,g) and 5(c) to find a parameter set that best describes the results. The fitting results are shown by the solid lines in each figure and the parameters extracted ($\theta_{SH}$, $\theta_{AH}$, $\lambda_N$, $\theta_{SN}$, $\theta_{AN}$, Re[$G_{MIX}$], Im[$G_{MIX}$]) are summarized in Table 1. (See the methods section for the details of the fitting process.) The spin Hall angle ($\theta_{SH}$) estimated for Ta and W underlayers are consistent with previous reports(*2, 6, 24, 25, 31*). These results show that the model can account for all results shown in Figs. 1-5 using a single set of parameters listed in Table 1. Note that the spin mixing conductance obtained from the fitting is mostly consistent with previous reports (see the Materials and methods section for the details).

To illustrate the effect of the spin Nernst effect on the transport properties more clearly, the spin Nernst and anomalous Nernst coefficients are numerically calculated using Eqs. (5) and (6) with three different spin Nernst angles, $\theta_{SN}=-\theta_{SH}$, $\theta_{SN}=0$, $\theta_{SN}=\theta_{SH}$. The open circles in Fig. 5(c) represent the scaled spin Hall magnetoresistance $SR_{SMR}$ calculated using the results of Figs. 1(d) and 3(e). As described above, $SR_{SMR}$ lies on the $\theta_{SN}=0$ line. This demonstrates that the internal electric field $\boldsymbol{E}_{INT}$ partly contributes to the spin current generation. In contrast, the spin Nernst coefficient $S_{SNE}$ (solid circles) lies closer to the $\theta_{SN}=-\theta_{SH}$ line. When the signs of $\theta_{SN}$ and $\theta_{SH}$ are opposite, contribution from the heat current induced spin current adds constructively to the $\boldsymbol{E}_{INT}$ induced spin current. Note that for the Ta underlayer films, the expected spin Nernst coefficient using Eq. (5) and the parameters defined in Table 1 (with $\theta_{SN}=-\theta_{SH}$) is ~0.01 (µV/K): this is smaller than the experimental resolution and we consider this is the reason we find no characteristic feature in the voltage measurements (Fig. 5(a)).

Furthermore, we show that the spin Nernst angle $\theta_{SN}$ can be extracted just from the experimental results. From Eqs. (1) and (5), we obtain:

$$\frac{\theta_{SN}}{\theta_{SH}} = \frac{S}{S_N}\left[\frac{S_{SNE}}{SR_{SMR}}+1\right] \quad (7)$$

In Fig. 5(d) we plot $\theta_{SN}/\theta_{SH}$ obtained by calculating $S_{SNE}/(SR_{SMR})$ using the results of Figs. 1(d), 3(e) and 5(c) (and Eq. (7)). The plot clearly shows the signs of the spin Nernst and spin Hall angles are opposite and the magnitude of the former is somewhat smaller than the latter. (Meyer *et al* have studied the spin Nernst effect in Pt/YIG and found that the signs of two angles are opposite for Pt too(*49*); however, the spin Nernst angle of Pt was reported to be larger than its spin Hall angle.) From numerical calculations, we find that $\theta_{SN}/\theta_{SH}$ is not susceptible to the values of the spin mixing conductance and the degree of longitudinal spin absorption (i.e. the spin polarization of the FM layer), which influences the absolute values of $R_{SMR}$ and $S_{SNE}$(*31*). The calculations also show that $\theta_{SN}/\theta_{SH}$ is not significantly influenced by contribution(s) from the anomalous Hall/anomalous Nernst effects and the spin Hall/spin Nernst effects, if any, of the FM layer as long as the HM layer thickness $d_N$ is larger than $\lambda_N$ (details will be reported elsewhere). When $d_N$ is smaller than $\lambda_N$, these effects can influence the value of $\theta_{SN}/\theta_{SH}$: the slight increase in $\theta_{SN}/\theta_{SH}$ found in Fig. 5(c) may be due to this contribution. We thus consider the large $d_N$ limit of $\theta_{SN}/\theta_{SH}$ provides a better estimate, from which we find $\theta_{SN}/\theta_{SH} \sim -0.7$.

**DISCUSSION**

Interestingly, the anomalous Nernst and the anomalous Hall angles(*50-54*) of CoFeB also possess the opposite sign (see Table 1), which results in a larger anomalous Nernst effect than otherwise. Theoretically, the sign of the Nernst and Hall angles do not necessarily have to match(*33*) as the Nernst angle ($\theta_{AN}$, $\theta_{SN}$) is defined by the energy derivative of the corresponding Hall conductivity near the Fermi energy, which can be positive or negative regardless of the sign of the Hall angle ($\theta_{AH}$, $\theta_{SH}$). Thus the sign as well as the magnitude of the Nernst angle can be very different from the Hall angle. The recently reported spin Hall tunneling spectroscopy(*55*) and/or the temperature gradient induced magnetization measurements(*56*) may provide access to information on the energy level dependence of the Hall conductivity and can be used to verify the relationship between the Hall and Nernst angles.

We briefly discuss contributions from other effects that may influence the signal due to the spin Nernst effect (see Supplementary Material Table S1 for more details). It has been reported that an unintended out of plane temperature gradient may develop during the application of an in-plane temperature gradient(*15, 43-45*). Under such circumstance, the anomalous Nernst effect of the FM layer can contaminate the signals observed in the voltage measurements. We observe such longitudinal voltage ($V_{XX}$) in film structures without the HM (W) layer and thicker FM (CoFeB) layer under application of $H_Y$. However, the $H_Y$ dependence of $V_{XX}$ is distinct: the values of $V_{XX}$ when the magnetization points along $+y$ and $-y$ are different for the anomalous Nernst voltage caused by the unintended out of plane temperature gradient (Supplementary material, Figs. S3(l-n)) whereas the values lie at the same level for the spin Nernst coefficient induced by the in-plane temperature gradient (Fig. 5(b)). For similar reason, the combined effect of the spin Seebeck effect within the FM layer and the inverse spin Hall effect of the HM layer under an out of plane temperature gradient can be excluded. The size of the unintended out of plane temperature gradient scales with the thickness of the CoFeB layer and it is smaller than the detection limit for the 1 nm thick CoFeB layer used here (see Supplemental material, Figs. S3(g-j)). We have also confirmed that the spin Nernst coefficient $S_{SNE}$ is negligible for heterostructures without the W layer (e.g. in Sub.|1 CoFeB|2 MgO|1 Ta) (see Supplemental material, Figs. S3(k) and S3(l)).

The results presented here not only provide insights into thermoelectric generation of spin current in heavy metals with strong spin orbit coupling but also have important implications on expanding the search of materials that can generate spin current. The spin Nernst effect may be able to generate spin current from materials that is not possible with the spin Hall effect, for example, in systems where the density of states at the Fermi level is zero. Of particular interest are the two dimensional chalcogenides and the Weyl semi metals in which the Fermi level coincides with the Dirac point. The spin Nernst effect may thus broaden material research on spin current generation beyond the current reach of the spin Hall effect.

**MATERIALS AND METHODS**

**A. Sample preparation and measurements**

All films are deposited using magnetron sputtering on non-doped silicon substrates coated with ~100 nm thick thermal oxides ($SiO_X$). Films are post-annealed at ~300°C for 1 hour prior to the device patterning processes. Optical lithography and Ar ion etching are used to pattern the films into wires and Hall bars. Contact pads made of 5 Ta|100 Au (unit in nanometer) are formed by a liftoff process.

All measurements are performed at room temperature. Temperature gradient across the substrate is applied by placing a ceramic heater on one side of the substrate and a heat absorbing Cu block on the other side. The substrate is fixed to the heater/Cu block using a thermally conducting double sided tape made of Al. The temperature profile of the system is studied using an infrared camera with Si substrates coated with black body matt (the surface emissivity is calibrated). The camera is used to ensure that the temperature gradient across the substrate is uniform. Due to the necessity of this coating, the temperature profile of the device under investigation cannot be monitored in real time: once the sample is coated with the black body matt, it is difficult to perform the voltage measurements. As the temperature gradient across the substrate largely depends on the contact between the substrate and the heater/Cu block, we have checked its variation by placing the substrate to the setup multiple times and monitored the temperature profile using the infrared camera. The variation of the temperature gradient is ~±10% of the average value. The horizontal error bars in Figs. 3(b,c) and 4(c) reflect this variation. The vertical error bars in the same figures represent the distribution of the voltage when measurements are repeated multiple times under the same contact between the substrate and heater/Cu block. The vertical error bars are smaller than the symbols, suggesting that the measurements are stable and the temperature gradient do not evolve once the substrate is fixed. Thus the dominant source of the measurement error originates from the uncertainty in the actual value of the temperature gradient across the substrate: the error bars in Figs. 3(d,e), 4(d-g) and 5(c,d) reflect this uncertainty.

**B. Fitting procedure**

Experimental results are fitted using Eqs. (1,2,5,6). Before carrying out the fitting, we determine the following parameters from the experimental results. The resistivity ($\rho_N$) of the HM layers are obtained by the $d_N$ dependence of $G_{XX}$ shown in Figs. 1(a,b). For the resistivity ($\rho_F$) of the FM (CoFeB) layer, we use a value from our previous studies(*57*). The Seebeck coefficients of the HM layer ($S_N$) and the FM layer ($S_F$) are estimated from the results presented in Figs. 3(d,e). We have also measured the Seebeck coefficient of the FM layer ($S_F$) independently using a film stack that does not include any HM layer: results are shown in Fig. S4(i). We find that $S_F$ estimated from film stacks with and without the HM layer are similar. The anomalous Hall angle ($\theta_{AH}$) and the anomalous Nernst angle ($\theta_{AN}$) of the FM (CoFeB) layer can be estimated by the zero HM thickness limit of the normalized $R_{AHE}$ (Figs. 2(c,d)) and the normalized $S_{ANE}$ (Figs. 4(f,g)), respectively.

We first fit $R_{SMR}$ (Figs. 1(c,d)) and $R_{AHE}$ (Figs. 2(c,d)) using Eqs. (1) and (2) to determine $\theta_{SH}$, $\lambda_N$, Re[$G_{MIX}$] and Im[$G_{MIX}$]. Note that in many previous studies, a transparent interface (Re[$G_{MIX}$]≫Im[$G_{MIX}$] and Re[$G_{MIX}$]≫$1/(2\rho_N\lambda_N)$) has been assumed to estimate the lower bound of $\theta_{SH}$. In such case, $G_{MIX}$ drops off from Eq. (1) and simplifies the fitting. Here we use Re[$G_{MIX}$] and Im[$G_{MIX}$] as the fitting parameters to account for the $d_N$ dependence of $R_{SMR}$ and $R_{AHE}$. For both underlayer films, we find that Im[$G_{MIX}$] has to be negative and larger in magnitude than Re[$G_{MIX}$]. Such characteristic $G_{MIX}$ is in agreement with the current induced torque found in similar heterostructures(*58-60*) according to the relation of $G_{MIX}$ and the torque(*61*). For the Ta underlayer films, the change in $R_{AHE}$ with $d_N$ is larger than what is expected from Eq. (2). As the anomalous Hall effect is known to be susceptible to interface states(*62, 63*), we infer that there are other effects that are not captured by Eq. (2).

Using these numbers ($\theta_{SH}$, $\lambda_N$, Re[$G_{MIX}$] and Im[$G_{MIX}$]), $S_{SNE}$ (Fig. 5(c)) and $S_{ANE}$ (Figs. 4(f,g)) are calculated using Eqs. (5) and (6) with $\theta_{SN}$ denoted in the legend of each figure.

**SUPPLEMENTARY MATERIALS**

**Supplementary material for this article is available.**

**Fig. S1. Magnetic properties of HM/CoFeB/MgO heterostructures.**

**Fig. S2. Spin Nernst magnetoresistance of Ta and W underlayer films.**

**Fig. S3. Thermoelectric properties of CoFeB thin films without HM layer.**

**Fig. S4. Comparison of parameters with and without the HM layer.**

**Table S1. Influence of other phenomena on the temperature gradient induced voltage measurements.**

**Acknowledgements:**

The authors thank S. Bosu and S. S-L. Zhang for fruitful discussions. This work was partly supported by JSPS Grant-in-Aid (15H05702), Casio Foundation, MEXT R & D Next-Generation Information Technology and the Spintronics Research Network of Japan. YCL is an International Research Fellow of the Japan Society for the Promotion of Science.

**Author Contributions**

M.H. and Y.S. planned the study. P.S and Y.L. carried out microfabrication, P.S., Y.L. and Y.S. measured the samples and analyzed the results with help of S.M. and M.H. S.T. developed the drift diffusion model. All authors discussed the data and commented on the manuscript.

**Competing interests**

The authors declare that they have no competing interests.

**Data and materials availability**

All data needed to evaluate the conclusions in the paper are present in the paper and/or the Supplementary Materials. Additional data related to this paper may be requested from the authors.

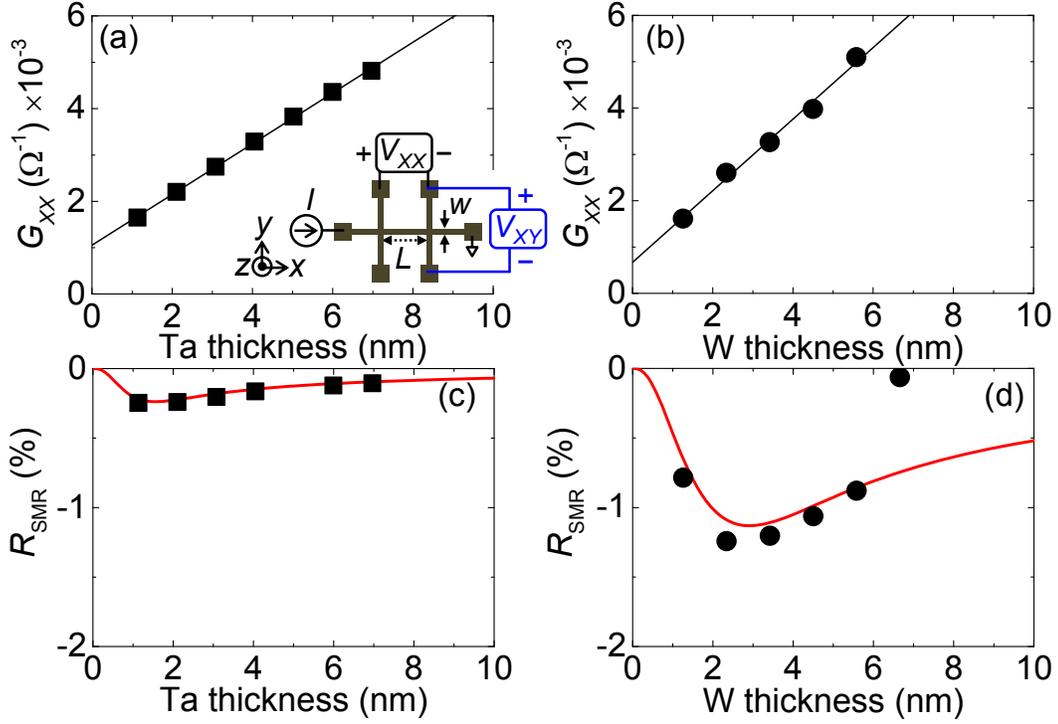

**Fig. 1. Longitudinal resistance and SMR of HM|CoFeB|MgO heterostructures.** (a,b) Sheet conductance $G_{XX}=L/(wR_{XX})$ vs. HM layer thickness $d_N$ for the Ta (a) and W (b) underlayer films. The solid lines show linear fit to the data in appropriate range of $d_N$. Schematic of the measurement setup is illustrated in the inset of (a). The inset of (b) is the expanded y-axis plot of the main panel. (c,d) Spin Hall magnetoresistance $R_{SMR}=\Delta R_{XX}/R_{XX}^z$ plotted against $d_N$ for the Ta (c) and W (d) underlayer films. The red solid lines are fit to the data using Eq. (1). Parameters used in the fitting are summarized in Table 1.

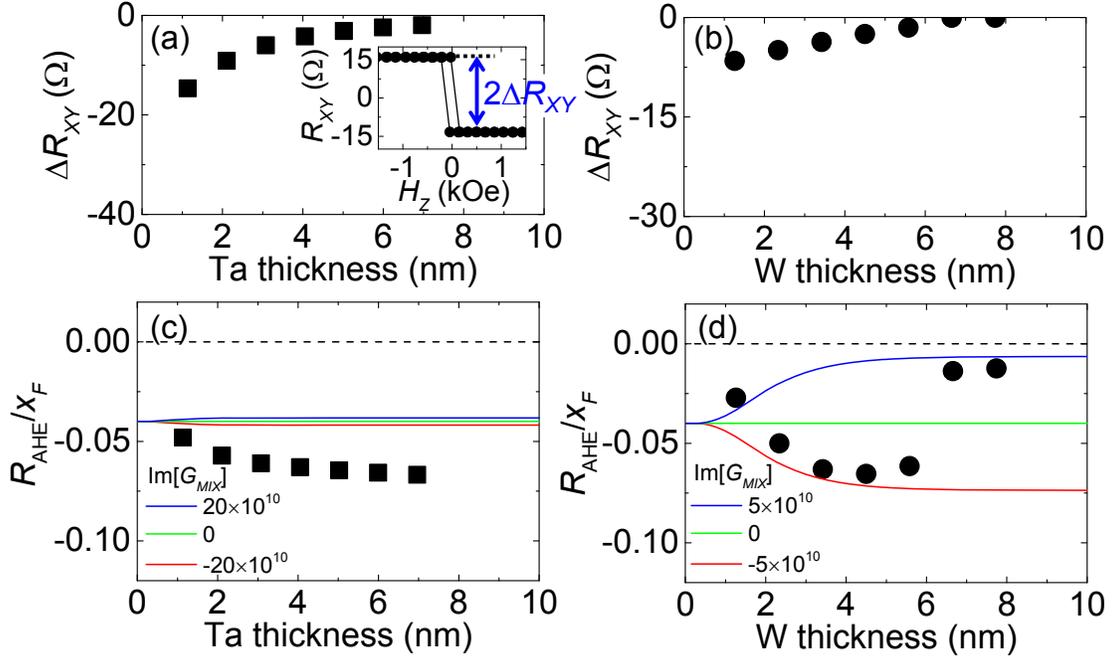

**Fig. 2. HM layer thickness dependence of the anomalous Hall resistance.** (a,b) The HM layer thickness $d_N$ dependence of the anomalous Hall resistance $\Delta R_{XY}$ for the Ta (a) and W (b) underlayer films. The inset of (a) shows $R_{XY}$ vs. $H_Z$ for sub.|~1.1 Ta|1 CoFeB|2 MgO|1 Ta (thickness in nm). Definition of $\Delta R_{XY}$ is schematically illustrated. (c,d) The normalized anomalous Hall coefficient $R_{AHE}/x_F = (\Delta R_{XY} L)/(w R_{XX}^Z x_F)$ plotted against $d_N$ for Ta (c) and W (d) underlayer films. The solid lines show fit to the data using Eq. (2) with three different values of Im[$G_{MIX}$]. Parameters used in the fitting are summarized in Table 1 except for Im[$G_{MIX}$] noted in the legend.

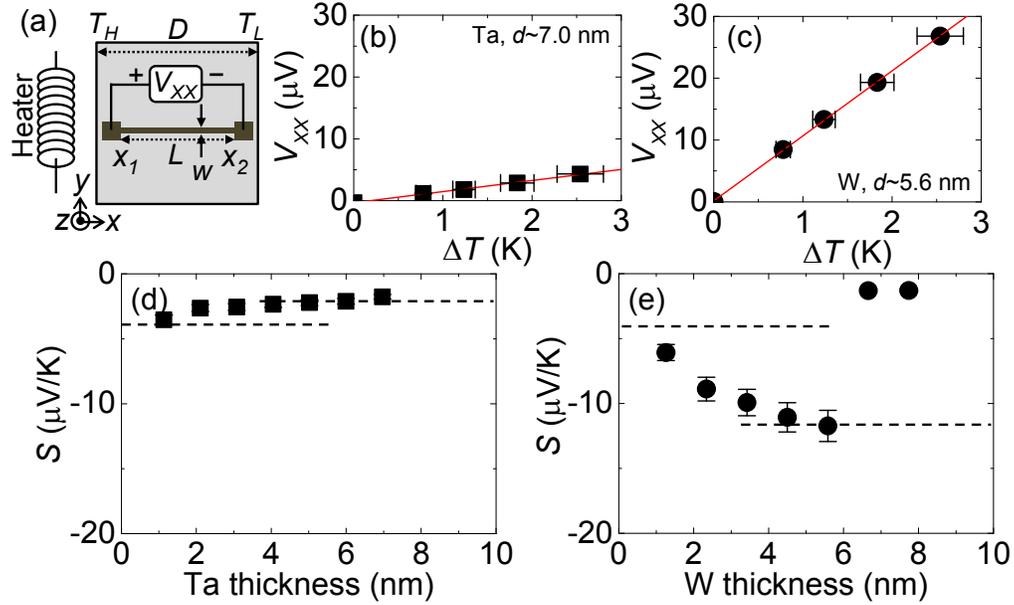

**Fig. 3. Seebeck coefficient of HM|CoFeB|MgO heterostructures.** (a) Schematic illustration of the measurement setup for temperature gradient induced longitudinal voltage. The bright square represents part of the substrate; the dark region indicates the area where the device is located. $D=0.7$ cm, $L=0.6$ cm, $w=50$ μm. (b,c) The longitudinal (Seebeck) voltage $V_{XX}$ measured as a function of the temperature difference $\Delta T$ for sub.|~7.0 Ta|1 CoFeB|2 MgO|1 Ta (b) and sub.|~5.6 W|1 CoFeB|2 MgO|1 Ta (c). The horizontal and vertical error bars represent, respectively, the uncertainty of the temperature gradient and the variation of the voltage under a fixed temperature gradient. (d,e) The Seebeck coefficient $S=-(V_{XX}/L)/(\Delta T/D)$ plotted against $d_N$ for Ta (d) and W (e) underlayer films. The error bars denote the variation of $S$ due to the uncertainty of the temperature gradient. The horizontal dashed lines are guide to the eye which provide estimate of the Seebeck coefficient of the HM and FM layers.

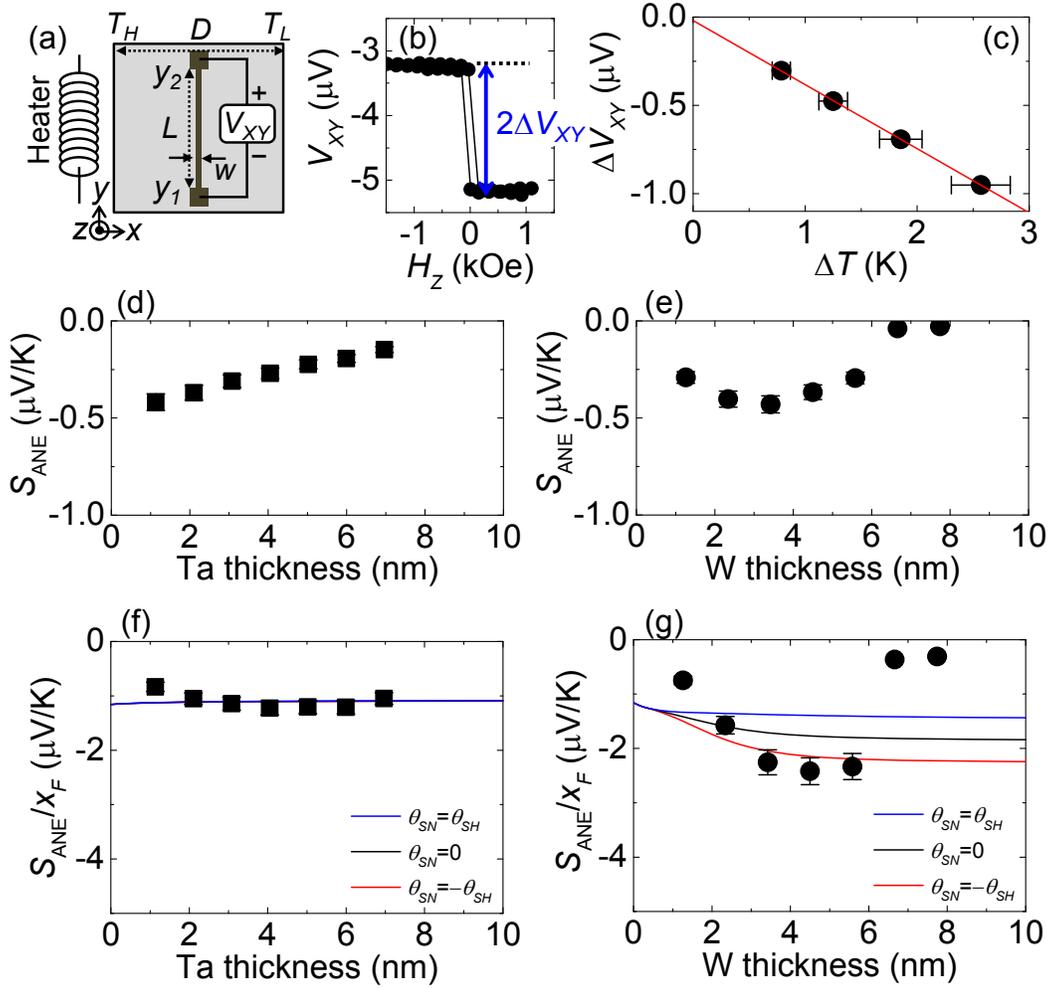

**Fig. 4. HM layer thickness dependence of the anomalous Nernst effect.** (a) Schematic illustration of the measurement setup for temperature gradient induced transverse voltage. The bright square represents part of the substrate; the dark region indicates the area where the device is located. $D$=0.7 cm, $L$=0.6 cm, $w$=50 μm. (b) The transverse voltage $V_{XY}$ vs. $H_z$ of sub.|~3.4 W|1 CoFeB|2 MgO|1 Ta when a temperature difference of $\Delta T$~2.5 K is applied. The definition of $\Delta V_{XY}$ is schematically drawn by the blue arrow. (c) $\Delta T$ dependence of the anomalous Nernst voltage $\Delta V_{XY}$ for the same sample described in (b). The horizontal and vertical error bars represent, respectively, the uncertainty of the temperature gradient and the variation of the voltage under a fixed temperature gradient. The red solid line shows linear fit to the data. (d,e) Anomalous Nernst coefficient $S_{ANE}=(\Delta V_{XY}/L)/(\Delta T/D)$ plotted against $d_N$ for the Ta (d) and W (e) underlayer films. (f,g) $d_N$ dependence of the normalized anomalous Nernst coefficient $S_{ANE}/x_F=(\Delta V_{XY}D)/(L\Delta Tx_F)$ for the Ta (f) and W (g) underlayer films. The error bars in (d-g) denote the variation of quantities due to the uncertainty of the temperature gradient. The solid lines in (f,g) show fit to the data using Eq. (6) with three different values of $\theta_{SN}$. Parameters used in the fitting are summarized in Table 1.

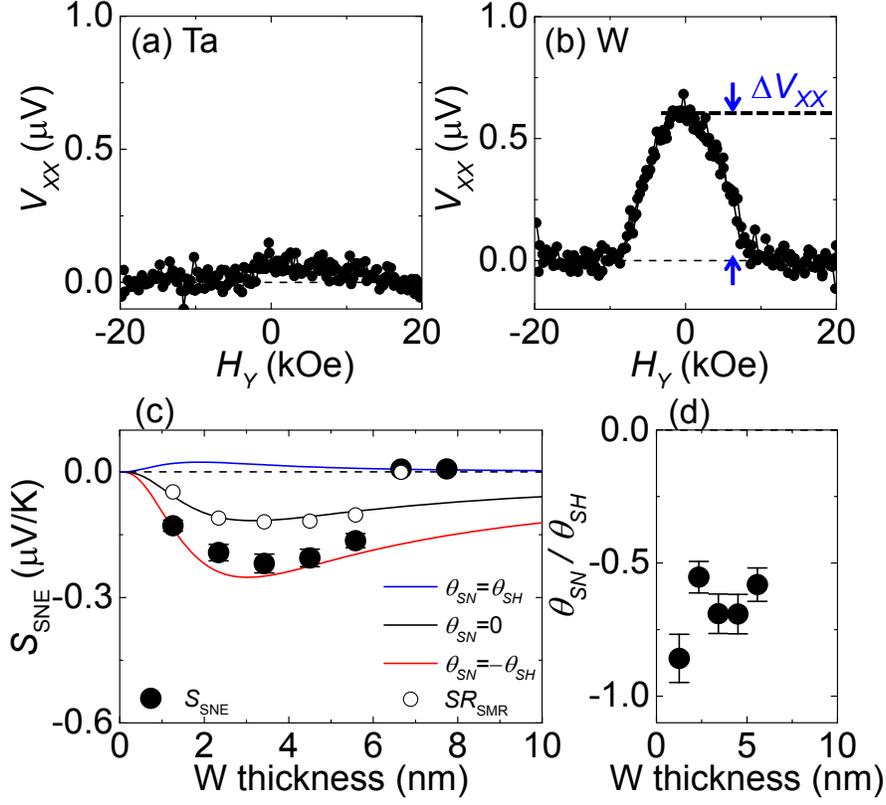

**Fig. 5. Signatures of the spin Nernst magnetoresistance.** (a,b) The longitudinal (Seebeck) voltage $V_{XX}$ vs. $H_Y$ for sub.|~3.1 Ta|1 CoFeB|2 MgO|1 Ta (a) and sub.|~3.4 W|1 CoFeB|2 MgO|1 Ta (b) when a temperature difference of $\Delta T$~3.5 K is applied. The definition of $\Delta V_{XX}$ is schematically drawn. (c) $d_N$ dependence of spin Nernst magnetoresistance $S_{SNE}=(\Delta V_{XX}/L)/(\Delta T/D)$ (solid circles) and the scaled spin Hall magnetoresistance $SR_{SMR}$ (open circles) for the W underlayer films. The solid lines show calculated $S_{SNE}$ using Eq. (5) with three different values of $\theta_{SN}$. Parameters used in the fitting are summarized in Table 1. (d) $d_N$ dependence of $\theta_{SN}/\theta_{SH}$ obtained from $S_{SNE}/(SR_{SMR})$ and the relation described in Eq. (7). The error bars in (c) and (d) denote the variation of quantities due to the uncertainty of the temperature gradient.

**Table 1. Parameters used to describe the experimental results.** Resistivity ($\rho_N$), Seebeck coefficient ($S_N$), spin diffusion length ($\lambda_N$), spin Hall angle ($\theta_{SH}$), spin Nernst angle ($\theta_{SN}$) of the heavy metal (HM) layer, and resistivity ($\rho_F$), Seebeck coefficient ($S_F$), anomalous Hall angle ($\theta_{AH}$), anomalous Nernst angle ($\theta_{AN}$) of the ferromagnetic metal (FM) layer in HM/FM/MgO heterostructure. Re[$G_{MIX}$] and Im[$G_{MIX}$] represent the real and imaginary parts of the spin mixing conductance $G_{MIX}$ at the HM/FM interface.

| Film structure | HM layer | | | | | F layer | | | | Interface | |
|---|---|---|---|---|---|---|---|---|---|---|---|
| | $\rho_N$ ($\mu\Omega$cm) | $S_N$ ($\mu$V/K) | $\lambda_N$ (nm) | $\theta_{SH}$ | $\theta_{SN}$ | $\rho_F$ ($\mu\Omega$cm) | $S_F$ ($\mu$V/K) | $\theta_{AH}$ | $\theta_{AN}$ | Re[$G_{MIX}$] ($\Omega^{-1}$cm$^{-2}$) | Im[$G_{MIX}$] ($\Omega^{-1}$cm$^{-2}$) |
| Ta/CoFeB | 183 | -2 | 0.5 | -0.13 | N/A | 160 | -4 | 0.04 | -0.25 | $2 \times 10^{10}$ | $-20 \times 10^{10}$ |
| W/CoFeB | 130 | -12 | 1.1 | -0.28 | varied | 160 | -4 | 0.04 | -0.25 | $2 \times 10^{10}$ | $-5 \times 10^{10}$ |

# Supplementary Material

# The spin Nernst effect in Tungsten

Peng Sheng[1], Yuya Sakuraba[1], Yong-Chang Lau[1,2], Saburo Takahashi[3], Seiji Mitani[1] and Masamitsu Hayashi[1,2*]

[1]*National Institute for Materials Science, Tsukuba 305-0047, Japan*
[2]*Department of Physics, The University of Tokyo, Bunkyo, Tokyo 113-0033, Japan*
[3]*Institute for Materials Research, Tohoku University, Sendai 980-8577, Japan*

## 1. Additional experimental results

*Magnetic properties of the heterostructures*

The magnetic properties of the heterostructures are evaluated using vibrating sample magnetometry. The results are presented in Fig. S1.

*Plots of Spin Nernst magnetoresistance*

The temperature gradient induced longitudinal voltage $V_{XX}$ is plotted as a function of magnetic field applied along the *y*-axis in Figs. S2(a-i) for the Ta (a,b) and W (c-i) underlayer films. $V_{XX}$ vs. $H_X$ and $H_Z$ for the W underlayer films are plotted in Figs. S2(j-m). The thickness of the HM layer (Ta or W) is listed in the legend. The applied temperature gradient $\Delta T$ is ~3.5 K.

*Thermoelectric properties of CoFeB*

To study contribution from the CoFeB layer on the spin transport properties of the Sub./HM/CoFeB/MgO/cap heterostructures, films without the heavy metal (HM) layer, i.e. Sub./CoFeB/MgO/cap heterostructures are studied. From hereafter, we refer to the latter heterostructure as "CoFeB films". The results are summarized in Figs. S3 and S4.

The CoFeB thickness dependence of the sheet conductance ($G_{XX}$) is plotted in Fig. S4(a). The resistance increases abruptly as the thickness ($t_F$) of the CoFeB layer is reduced to near ~1 nm. Below $t_F$~1 nm, the device resistance cannot be evaluated. Note that for film structures with the HM underlayer, the resistance of devices with $t_F$~1 nm is significantly smaller. We consider such difference arises due to change in the film growth mode. The HM layer serves as a good seed layer for growing films on SiOx surfaces; Ta is a good example that is well known and widely used. Without the HM layer, the growth mode of CoFeB changes from layer growth to island growth. For thin CoFeB films ($t_F \lesssim$ 1 nm) without the HM layer, the film morphology may not be uniform and continuous. For these reasons, studies of the CoFeB films without the HM underlayer are limited to thicknesses ($t_F$) larger than ~1 nm.

From the in-plane field ($H_Y$) dependence of the longitudinal resistance ($R_{XX}$), we find the spin Hall magnetoresistance (SMR) of the CoFeB films to be ~-0.01% to ~0.1% (Figs. S3a-d). The CoFeB thickness dependence of the SMR ($R_{SMR}=\Delta R_{XX}/R_{XX}^Z$) is plotted in Fig. S4(b). Since the signal is small, it is not clear what is causing the positive SMR for the thinner CoFeB films: the resistance measurements may capture effects other than SMR (e.g. anisotropic magnetoresistance related to interface states). However, the magnitude of the SMR found in the CoFeB films is one order of magnitude smaller than that of Ta underlayer films and nearly two orders of magnitude smaller than that of the W underlayer films.

The out of plane field ($H_Z$) dependence of the anomalous Hall resistance and the anomalous Nernst voltage are plotted in Figs. S3(e-j). The normalized anomalous Hall coefficient $R_{AHE}/x_F=(\Delta R_{XY}L)/(R_{XX}^Z w x_F)$ and the normalized anomalous Nernst coefficient $S_{ANE}/x_F=(\Delta V_{XY}D)/(\Delta T L x_F)$ of the CoFeB films are plotted as a function of its thickness ($t_F$) in Figs. S4(c) and S4(f), respectively. These results are compared to the results of heterostructures with the HM layer (Figs. S4(d,e) and S4(g,h)). We find that the small HM layer thickness ($d_N$) limit of $R_{AHE}/x_F$ and $S_{ANE}/x_F$ more or less agree with that of the thick CoFeB films. As the thicker CoFeB films are likely more uniform than its thinner counterpart, it may be reasonable to compare results from the thicker CoFeB films with those of the small $d_N$ limit of films with the HM underlayer. Note that when fitting $S_{ANE}/x_F$ with Eq. (6), we use a value of $\theta_{AN}$ that results in a larger $S_{ANE}/x_F$ than that of the CoFeB films. In order to describe all results ($R_{SMR}$, $R_{AHE}/x_F$, $S_{ANE}/x_F$ and $S_{SNE}$) with a single parameter set, $\theta_{AN}$ needs to take a value slightly larger than what is expected for CoFeB films without the HM layer.

The Seebeck coefficient $S=-(V_{XX}/L)/(\Delta T/D)$ of the CoFeB films are compared to that of the HM layer included heterostructures in Fig. S4(i-k). The Seebeck coefficient of the CoFeB films shows little dependence on the CoFeB thickness (Fig. S4(i)). The small $d_N$ limit of $S$ is consistent with that of the CoFeB films.

Finally, the temperature gradient induced longitudinal voltage ($V_{XX}$) of the CoFeB films is plotted as a function of $H_Y$ in Figs. S3(k-n). For the thicker CoFeB films, we find signals which resemble that of the anomalous Nernst effect (ANE), i.e. the level of $V_{XX}$ is different for positive and negative $H_Y$. It should be noted that the CoFeB films used here has easy plane anisotropy, different from the heterostructures with the HM underlayer that possess uniaxial perpendicular magnetic anisotropy. We consider the ANE like signal appears due to an unintended out of plane temperature gradient applied across the film. Influence of such out of plane temperature gradient on the thermoelectric measurements has been pointed by previous reports(*15, 43-45*). However, it should be noted that the ANE-like signal decreases with decreasing CoFeB thickness, consistent with previous report(*43*).

As the CoFeB layer thickness is reduced, the noise level of $V_{XX}$ tends to increase (see Figs. S3(k-n)). We consider such increase in the noise level is related to the film morphology and the resistance of the CoFeB layer: thinner CoFeB layers are less likely to form a continuous layer and thus the resistance considerably increases, giving rise to large Johnson noise. However, in the thicker CoFeB films with smaller noise level (Figs. S3(k-n)), we find almost no signal that resembles that of the $H_Y$ dependence of $V_{XX}$ shown in Fig. 5(b).

*Magnitude of the unintended out of plane temperature gradient*

We can estimate the magnitude of the unintended out of plane temperature gradient found in the samples shown in Figs. S3(m) and S3(n) using the corresponding ANE measurements: see Figs. S3(i) and S3(j). Comparing the voltage difference when the magnetization direction is reversed, we estimate the out of plane temperature gradient to be ~15% (~23%) of the in-plane temperature gradient for the samples with CoFeB thickness of ~1.9 nm (~2.2 nm). The results presented in Figs. S3(k,l,m,n) show that the out of plane temperature gradient and consequently the ANE-like signal tend to decrease as the CoFeB layer thickness is reduced. This is in accordance with previous report, which concluded that an unintended out of plane temperature gradient increases with increasing thickness of the magnetic layer due to the difference in the thermal conductivity of the substrate and the film(*43*). As the thickness of the CoFeB layer ($t_F$~1 nm) used for the studies presented in Figs. 1-5 are thinner than those shown in Fig. S3 and S4, we consider the effect of the unintended out of plane temperature gradient on the longitudinal voltage in the W underlayer films is negligible.

**2. Discussion related to other effects that may influence the voltage measurements**

*Spin Seebeck effect*

Here we discuss the effect of the spin Seebeck effect(*9, 12*) on the voltage measurements. First we consider the influence of the unintended out of plane temperature gradient (along *z* in Figs. 3(a) and 4(a)).

For the anomalous Nernst effect measurements, the transverse voltage ($V_{XY}$) is measured against the out of plane field ($H_Z$). In this experimental setting, the magnetization of the FM layer always points along the film normal. With an unintended out of plane temperature gradient, the spin Seebeck effect within the FM layer will generate, if any, spin current with both spin ($\sigma$) and flow ($J_S$) pointing along the film normal (along *z*). The inverse spin Hall effect (ISHE) will not generate any voltage ($V_{XY}$) in the HM layer since the charge current $J$ is proportional to the cross product of $\sigma$ and $J_S$, i.e. $J \propto J_S \times \sigma$. Thus we expect no contribution from the ISHE on the transverse voltage measurements with the unintended out of plane temperature gradient (this condition corresponds to the cartoon of Table S1, row D, column "Setup 2").

Next, we consider the influence of the spin Seebeck effect and the ISHE on $V_{XX}$ (with the unintended out of plane temperature gradient). Note that for the spin Nernst coefficients measurements, we compare the difference of $V_{XX}$ when the magnetization of the FM layer is pointing along the film normal (*z*-axis) and along the film plane (*y*-axis). $V_{XX}$ is expected to be zero when the magnetization is pointing along the film normal since the directions of both spin and flow of the spin current, if any, are parallel ($J_S \parallel e_z$, $\sigma \parallel e_z$). When the magnetization is rotated toward the *y*-axis, there is a possibility that the spin Seebeck effect and the ISHE generate a longitudinal voltage. Under such circumstance, $H_Y$ will dictate the spin direction ($\sigma$) of the spin current and thus $V_{XX}$ changes its sign when the magnetization direction is reversed along the *y*-axis due to the ISHE (e.g. $J_S \parallel e_z, \sigma \parallel \pm e_y, J \propto J_S \times \sigma \sim \mp e_x$); see cartoon of Table S1, row D, column "Setup 1". In our experiments, we do not observe such difference in $V_{XX}$ when the magnetization is reversed between +*y* and −*y* (see Fig. 5(b) and Figs. S2(c-g)). We thus consider such contribution from the unintended out of plane temperature gradient, spin Seebeck effect and ISHE is negligible in the spin Nernst coefficient measurements.

With the in-plane temperature gradient (applied along *x* as in Figs. 3(a) and 4(a)), the spin Seebeck effect and the ISHE can generate a non-zero $V_{XX}$ at the hot and cold ends of the heterostructures under application of $H_Y$. However, since $V_{XX}$ at the hot and cold ends are opposite (e.g. due to the diffusion of accumulated spins, $\boldsymbol{J}_S \parallel \boldsymbol{e}_z$ for the hot end and $\boldsymbol{J}_S \parallel -\boldsymbol{e}_z$ for the cold end; $\boldsymbol{\sigma} \parallel \pm\boldsymbol{e}_y$ is determined by $H_Y$, thus $\boldsymbol{J} \propto \boldsymbol{J}_S \times \boldsymbol{\sigma} \sim \mp\boldsymbol{e}_x$ for the hot and $\boldsymbol{J} \sim \pm\boldsymbol{e}_x$ for the cold end), they will cancel out and result in a net zero $V_{XX}$ (Table S1, row C, column "Setup 1"). The transverse voltage $V_{XY}$ vs. $H_Z$ is also expected to be zero since the spin accumulation ($\boldsymbol{\sigma}$) due to the spin Seebeck effect at the edges will not generate any $V_{XY}$ due to the ISHE ($\boldsymbol{\sigma}$ and $\boldsymbol{J}_S$ are both along *z*). This is shown in Table S1, row C, column "Setup 2".

*Anomalous Nernst effect*

Here we consider the influence of the anomalous Nernst effect together with the unintended out of plane temperature gradient (along *z* in Figs. 3(a) and 4(a)). The transverse charge current due to the ANE is: $\boldsymbol{J} \propto \nabla T \times \boldsymbol{M}$. Under application of $H_Z$, $V_{XX}$ and $V_{XY}$ are expected to be zero since the magnetization (*M*) and the temperature gradient ($\nabla T$) are parallel. Thus $V_{XY}$ vs. $H_Z$ is expected to be flat, as shown in Table S1, row E, column "Setup 2".

The in-plane field ($H_Y$) dependence of $V_{XX}$ will show different voltage levels at $+H_Y$ and $-H_Y$ since *M* is determined by $H_Y$ (e.g. $\nabla T \parallel \boldsymbol{e}_z$, $\boldsymbol{M} \parallel \pm\boldsymbol{e}_y$, $\boldsymbol{J} \propto \nabla T \times \boldsymbol{M} \sim \mp\boldsymbol{e}_x$).

Table S1, row E, column "Setup 1" shows a cartoon that one would expect for this configuration.

In all cases, the expected curves are different from the experimental results shown in the main text. We thus consider the anomalous Nernst effect driven by an unintended out of plane temperature gradient has little influence on the measurements we performed.

*Spin Hall and spin Nernst effect of the FM layer*

The spin Hall and spin Nernst effects of the FM (CoFeB) layer contribute to the measured voltage in a similar way as the two effects of the HM layer do (see cartoons of Table S1, row F, columns "Setup 1" and "Setup 2"). Although the size of the spin Hall angle of ferromagnetic materials is an issue currently debated, reports up to date show that the effect is much smaller than that of the heavy metal layers(*46-48*). For example, the spin Hall angle of Py has been reported to be ~0.005-0.013 (refs. (*46, 47*)), significantly smaller than that of Ta (~0.13) and W (~0.28).

Contribution from the spin Hall/spin Nernst effects of the FM layer on the spin transport properties of the HM/FM bilayer can be included in the model and we find that the ratio $\theta_{SN}/\theta_{SH}$ is not significantly influenced with moderate values (smaller than ~0.1) of $\theta_{SN}$ and $\theta_{SH}$ of the FM layer. Details of the model calculations will be presented elsewhere. We thus consider obtaining $\theta_{SN}/\theta_{SH}$ from the experimental results (Figs. 1(d), 3(e), 5(c)) provides a robust way to estimate the spin Nernst effect of non-magnetic materials.

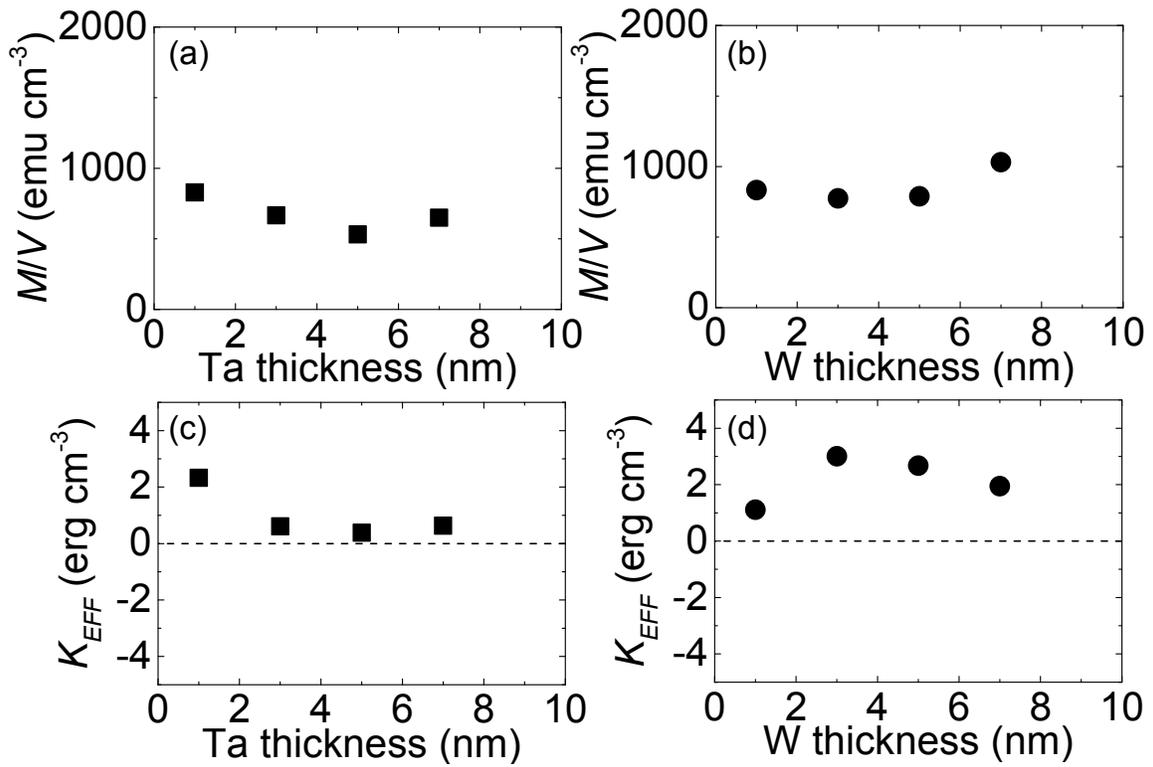

**Fig. S1. Magnetic properties of HM|CoFeB|MgO heterostructures.** (a-d) Saturated magnetic moments per unit volume $M/V$ (a,b) and effective magnetic anisotropy energy $K_{EFF}$ (c,d) plotted as a function of the HM layer thickness $d_N$ for Ta (a,c) and W (b,d) underlayer films.

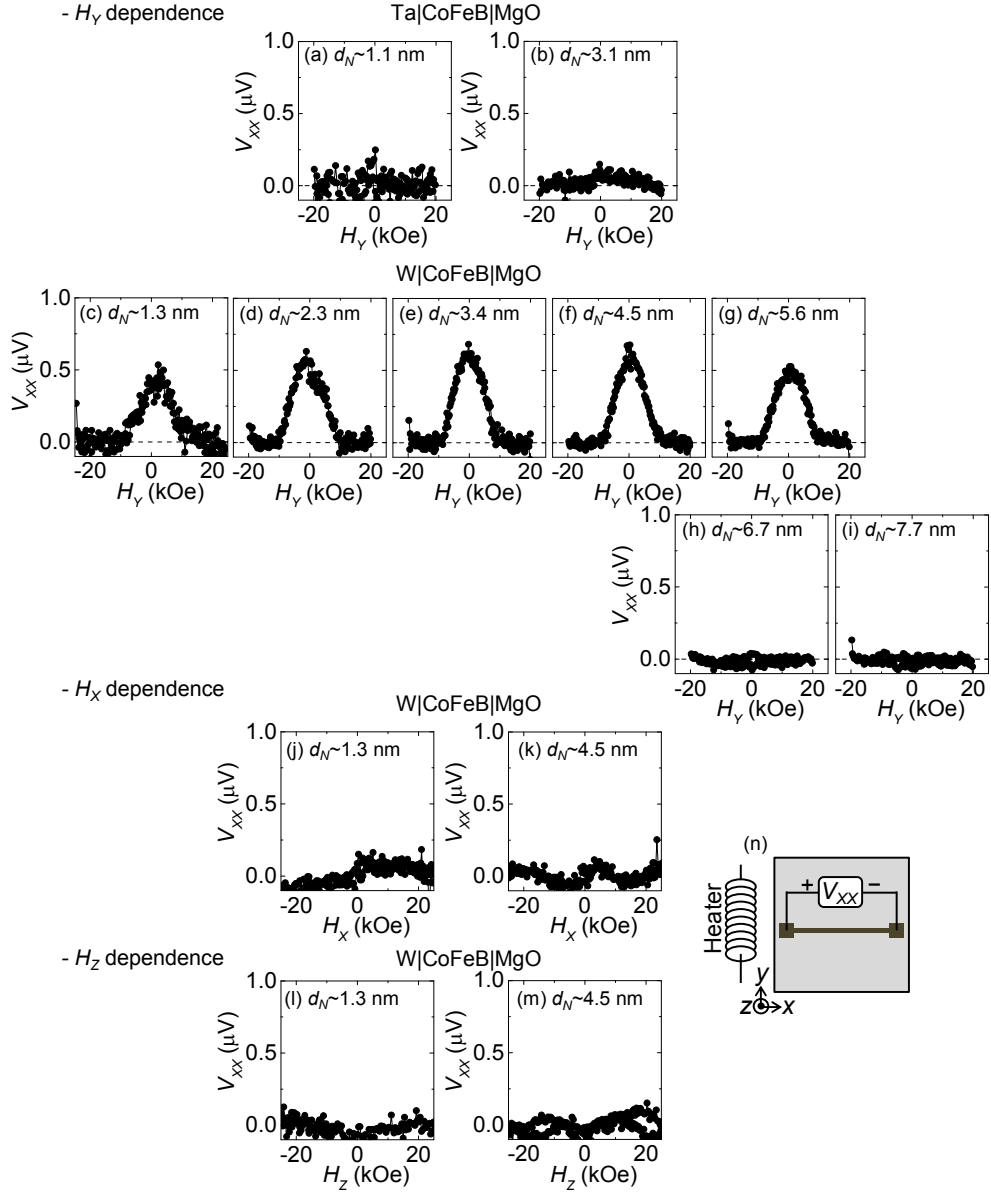

**Fig. S2. Spin Nernst magnetoresistance of Ta and W underlayer films.** (a-h) The longitudinal (Seebeck) voltage $V_{XX}$ vs. $H_Y$ of the Ta underlayer films (a,b) and the W underlayer films (c-i). (j-m) $H_X$ and $H_Z$ dependence of $V_{XX}$ for the W underlayer films. A temperature difference $\Delta T \sim 3.5$ K is applied across part of the substrate ($D \sim 7$ mm). $d_N$ denotes the thickness of the HM (Ta or W) underlayer. (n) Schematic of the experimental setup and the coordinate axis.

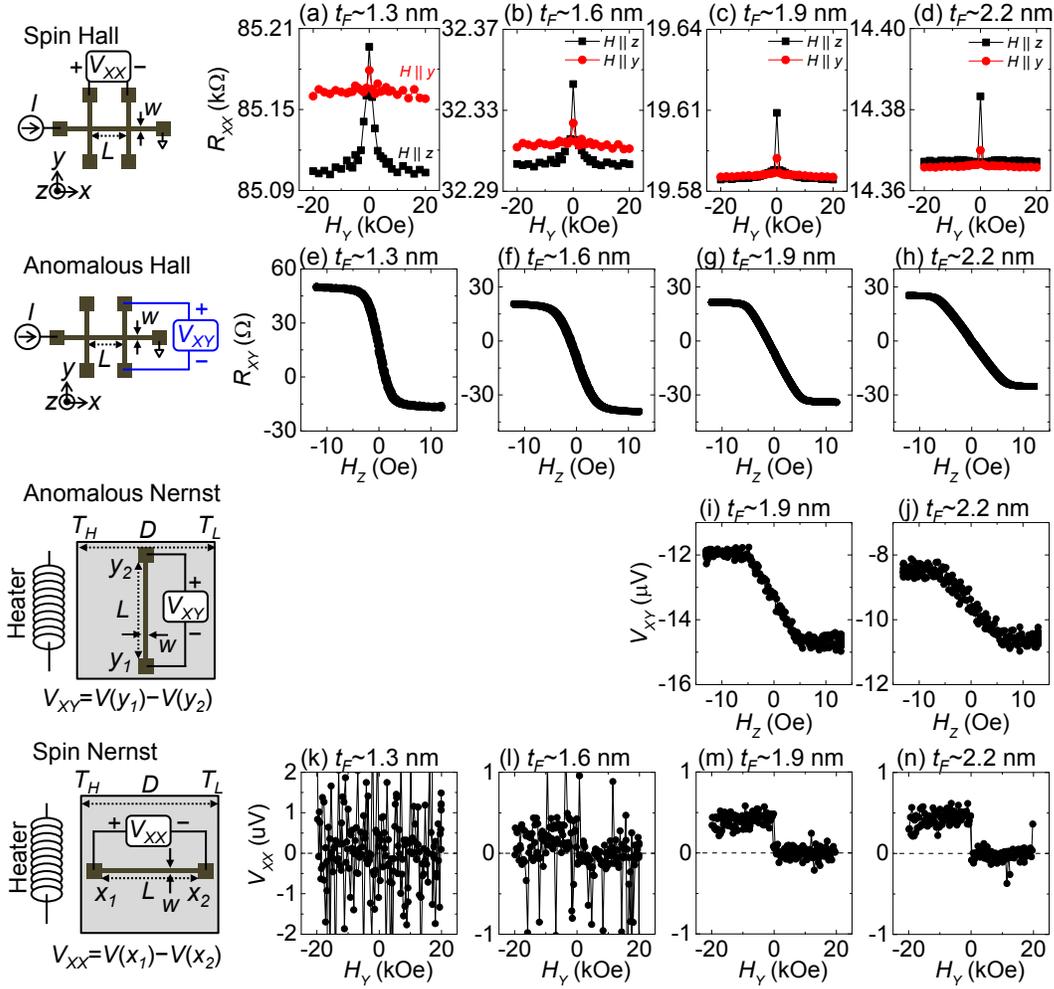

**Fig. S3. Thermoelectric properties of CoFeB thin films without the HM layer.** (a-n) In plane field ($H_Y$) dependence of the longitudinal resistance $R_{XX}$ (a-d), out of plane field ($H_Z$) dependence of the transverse resistance $R_{XY}$ (e-h), $H_Z$ dependence of the temperature gradient induced transverse voltage $V_{XY}$ (i,j) and $H_Y$ dependence of the temperature gradient induced longitudinal voltage $V_{XX}$ (k-n) for sub.|$t_F$ CoFeB|2 MgO|1 Ta (thickness in nm) heterostructures. The thickness of the CoFeB layer ($t_F$) is listed on top of each panel. Schematics of the experimental setup are displayed on the left. For (i-n), the applied temperature gradient $\Delta T$ is ~3.5 K. The longitudinal voltages $V_{XX}$ shown in (k-n) are vertically shifted so that the large $H_Y$ limit of $V_{XX}$ becomes zero.

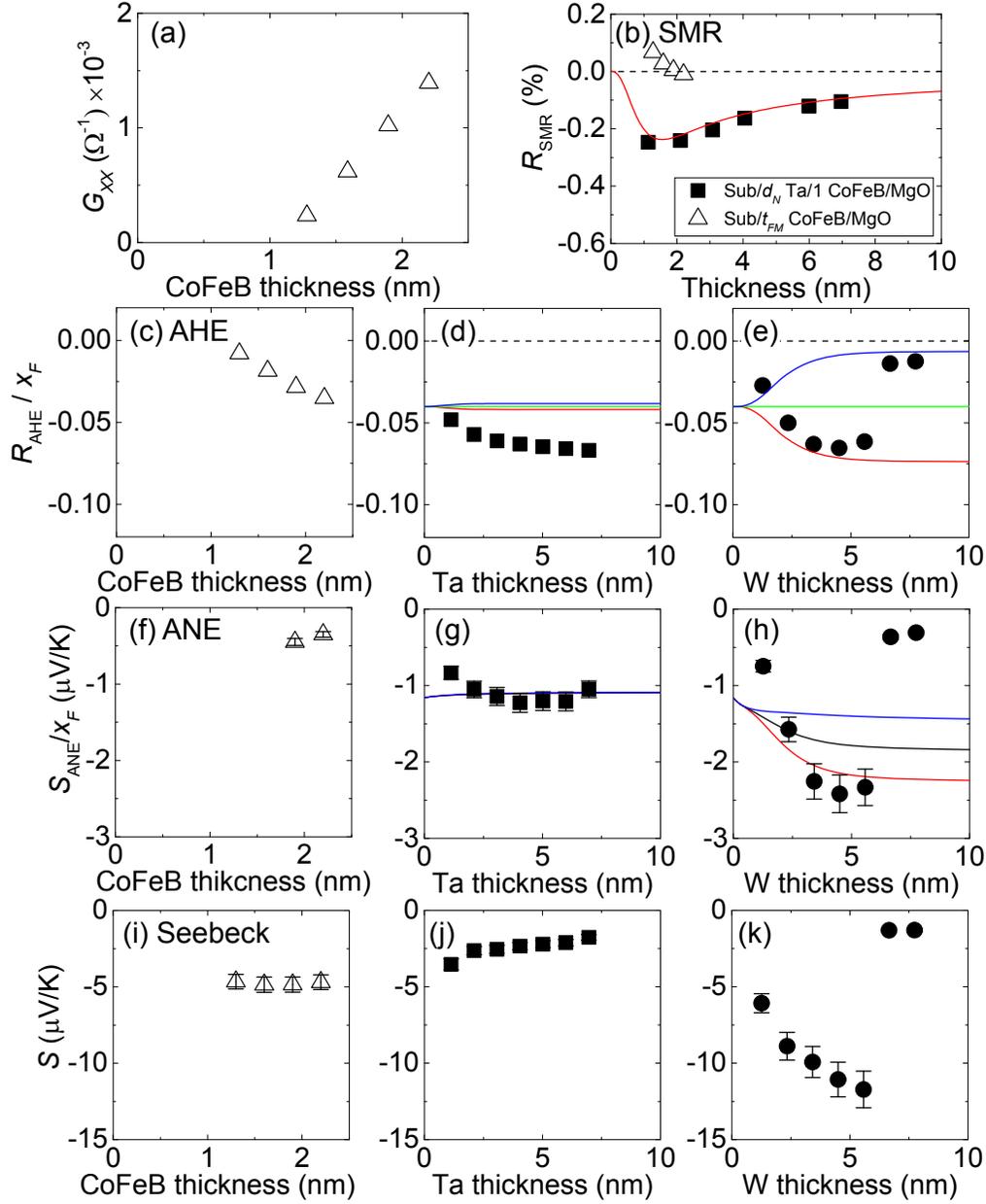

**Fig. S4. Comparison of parameters with and without the HM layer.** (a,b) CoFeB thickness dependence of the sheet conductance $G_{XX}=L/(wR_{XX})$ (a) and the spin Hall magnetoresistance $R_{SMR}=\Delta R_{XX}/R_{XX}^{z}$ (b). $R_{SMR}$ for the Ta underlayer films are shown as a reference. (c-e) Normalized anomalous Hall coefficient $R_{AHE}/x_F=(\Delta R_{XY}L)/(R_{XX}^{z}wx_F)$ plotted against CoFeB (c), Ta (d) and W (e) layer thicknesses. (f-h) CoFeB (f), Ta (g) and W (h) layer thicknesses dependence of the normalized anomalous Nernst coefficient $S_{ANE}/x_F=(\Delta V_{XY}D)/(\Delta TLx_F)$. (i-k) Seebeck coefficient $S=-(V_{XX}/L)/(\Delta T/D)$ plotted as a function of CoFeB (i), Ta (j) and W (k) layer thicknesses. The error bars in (f-k) denote variation of quantities due to the uncertainty of the temperature gradient. Film structure used are sub.|$t_F$ CoFeB|2 MgO|1 Ta (a,b,c,f,i), sub.|$d_N$ Ta|1 CoFeB|2 MgO|1 Ta (d,g,j) and sub.|$d_N$ W|1 CoFeB|2 MgO|1 Ta (e,h,k).

**Table S1. Influence of other phenomena on the temperature gradient induced voltage measurements.**[*]

| | | Setup 1<br>$V_{XX}$ vs. $H_Y$ (Fig.3a) | Setup 2<br>$V_{XY}$ vs. $H_Z$ (Fig.4a) |
|---|---|---|---|
| Magnetization | | 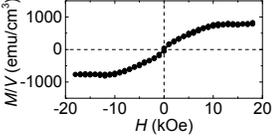 | 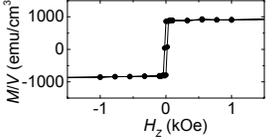 |
| Sources | | | |
| A | $\Delta T$ along $x$<br>**Spin Nernst (HM)**<br>**Spin Hall magnetoresistance** | 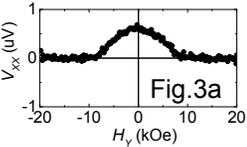Fig.3a | 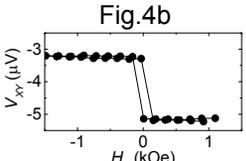Fig.4b |
| B | $\Delta T$ along $x$<br>**Anomalous Nernst (FM)** | 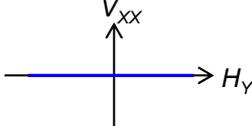 | |
| C | $\Delta T$ along $x$<br>Spin Seebeck effect (FM)<br>Inverse spin Hall effect (HM) | 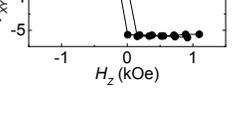 | 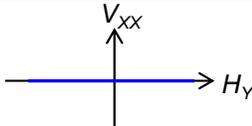 |
| D | (Unintended) $\Delta T$ along $z$<br>Spin Seebeck effect (FM)<br>Inverse spin Hall effect (HM) | 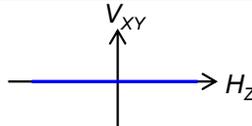 | 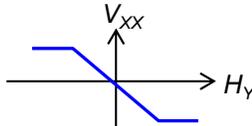 |
| E | (Unintended) $\Delta T$ along $z$<br>Anomalous Nernst (FM) | 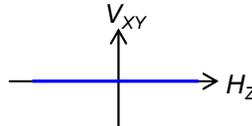 | 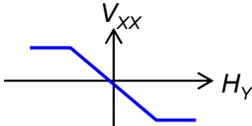 |
| F | $\Delta T$ along $x$<br>Spin Nernst (FM)<br>Spin Hall magnetoresistance | 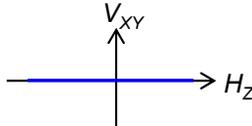 | 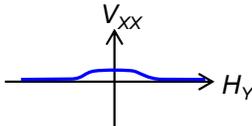 |

[*]The two columns ("Setup 1" and "Setup 2") show the in-plane field ($H_Y$) and out of plane field ($H_Z$) dependence of the properties represented by the y-axis title of each panel. The top row show measured $M$-$H$ loops of sub.|3 W|1 CoFeB|2 MgO|1 Ta. The rows labeled A-F illustrate the expected transverse ($V_{XY}$) and longitudinal ($V_{XX}$) voltages when the phenomena indicated in the corresponding left column take place. For example, if an unintended out of plane temperature gradient ($\Delta T$ along $z$) is applied and the spin Seebeck effect occurs in the FM layer together with the inverse spin Hall effect in the HM layer (row D), we expect an asymmetric $V_{XX}$ vs. $H_Y$ and a nearly zero $V_{XY}$ vs. $H_Z$. For rows A and B, $V_{XY}$ is due to both the spin Nernst and anomalous Nernst effects; the rows are thus merged.